\documentclass[twocolumn,preprintnumbers,elsart]{revtex4}
\usepackage{amssymb}
\usepackage{amsmath}
\usepackage{graphicx}
\usepackage{dcolumn}
\usepackage[center]{subfigure}

\begin{document}

\title{Semidiscrete quantum droplets and vortices}
\author{Xiliang Zhang$^{1,a}$, Xiaoxi Xu$^{1,a}$, Yiyin Zheng$^{1}$, Zhaopin
Chen$^{2}$, Bin Liu$^{1}$, Chunqing Huang$^{1}$}
\author{Boris A. Malomed$^{2,1}$}
\author{Yongyao Li$^{1,2}$}
\email{yongyaoli@gmail.com}
\affiliation{$^{1}$School of Physics and Optoelectronic Engineering, Foshan University,
Foshan 528000, China \\
$^{2}$ Department of Physical Electronics, School of Electrical Engineering,
Faculty of Engineering, and the Center for Light-Matter Interaction, Tel
Aviv University, Tel Aviv 69978, Israel.\\
$^{a}$These two authors contributed equally to this paper.}

\begin{abstract}
We consider a binary bosonic condensate with weak mean-field (MF) residual
repulsion, loaded in an array of nearly one-dimensional traps coupled by
transverse hopping. With the MF force balanced by the effectively
one-dimensional attraction, induced in each trap by the Lee-Hung-Yang
correction (produced by quantum fluctuations around the MF state), stable
onsite-centered and intersite-centered semidiscrete quantum droplets (QDs) emerge in
the array, as fundamental ones and self-trapped vortices, with winding
numbers, at least, up to $5$, in both tightly-bound and quasi-continuum
forms. The application of a relatively strong trapping potential leads to
squeezing transitions, which increase the number of sites in fundamental
QDs, and eventually replace vortex modes by fundamental or dipole ones. The
results provide the first realization of stable semidiscrete vortex QDs,
including ones with multiple vorticity.
\end{abstract}

\maketitle

\textit{Introduction and the model}. Recent works with binary Bose-Einstein
condensates (BECs) have led to a breakthrough in studies of quantum matter,
predicting and experimentally realizing ultradilute superfluids which form
quantum droplets (QDs). They were predicted in the three-dimensional (3D)
setting \cite{Petrov2015}, as well as in its 2D and 1D reductions \cite%
{Petrov2016}, on the basis of mean-field (MF) Gross-Pitaevskii equations
(GPEs) with the Lee-Hung-Yang (LHY) corrections, that account for quantum
fluctuations around MF states \cite{LHY1957}. In 3D and 2D geometries, the
LHY terms are repulsive, helping to stabilize a binary condensate against
the collapse driven by the cross-attraction between its components, which
slightly exceeds self-repulsion in each one, the residual attraction being
balanced by the LHY terms. As a result, stable multidimensional soliton-like
states may be created in the form of QDs, which is a problem of great
interest \cite{review}, a challenging issue being stability of 2D and 3D
solitons against the collapse. The prediction was followed by the creation
of quasi-2D \cite{Cabrera2017,Cheiney2018} and isotropic 3D \cite{Ing1,Ing2}
QDs in a binary condensate of two different states of $^{39}$K atoms. The
competition of long-range attractive interactions and LHY repulsion has also
made it possible to create stable QDs in single-component condensates of
dipolar atoms \cite{Pfau}-\cite{Saito2016}. In addition to their
significance to fundamental studies, QDs offer potential applications, such
as the design of matter-wave interferometers \cite{Tolra2016}. An essential
extension is the recent prediction of stable 3D \cite{3D-vortex} and 2D \cite%
{we2019} two-component QDs with embedded vorticity and robust
necklace-shaped clusters \cite{cluster} (vortex QDs in dipolar condensates
were found to be unstable \cite{Macri}).

The reduction of the MF system with the LHY corrections to the 1D
configuration (the condensate loaded in a cigar-shaped trap subject to
strong transverse confinement \cite{cigar,Luca,Hulet,ZouZeng}) changes the setting,
making the LHY term attractive, on the contrary to its repulsive sign in
higher dimensions \cite{Petrov2016}. Accordingly, the most relevant case is
one with the residual cubic MF\ repulsion (the inter-component attraction
being slightly weaker than the repulsion in each component) competing with a
quadratic term representing the LHY-induced attraction. Self-trapped states
in this model demonstrate Gaussian-like and flat-top shapes in the case of
relatively small or large numbers of atoms, respectively \cite%
{Astrakharchik2018}. The next natural step is the consideration of a
tunnel-coupled pair of 1D waveguides, in which spontaneous symmetry breaking
of QDs was predicted \cite{Bin2019} (similar systems, combining the LHY term
and linear mixing between two components, were introduced too \cite%
{SalasMacri,Sala}).

The availability of optical lattices (OLs) for BEC\ experiments \cite%
{Maciek,Bloch} suggests to consider a setting in the form of an array of
1D traps, coupled by hopping to adjacent ones. Similar configurations were
broadly considered in optics, in the form of parallel-coupled arrays of
fibers and stacks of planar waveguides, in temporal- and spatial-domain
forms, respectively \cite{Christodoul}-\cite{Blit}. In the combination with
intra-core nonlinearity, they give rise to 2D semidiscrete solitons, which
are continuous objects along the guiding cores and discrete in the
transverse direction \cite{Rub1,Rub2,Jena,Blit,Osgood}.

In this work, we aim to introduce semidiscrete QDs in the system of
transversely coupled 1D traps, filled by the binary condensate which
features the combination of the weak MF repulsion and LHY-induced attraction
in each trap. Subjects of special interest are semidiscrete solitary
vortices, which were not considered previously. We produce stable solutions
for both fundamental (zero-vorticity) and vortical semidiscrete QDs, with
the winding number up to $S=5$. In the 2D continuum form, bright vortex
solitons were produced in various models \cite{Quiroga}-\cite{Adhikari2},
\cite{3D-vortex,we2019}, the main issue being their stability \cite{PhysD}.
While the stabilization of vortices was theoretically elaborated in diverse
forms, it was demonstrated experimentally only in nonlocal media \cite%
{liq-cryst}. In settings with local nonlinearity, self-trapped vortices were
experimentally observed in transient forms \cite{Pertsch,Cid}. On the other
hand, vortex solitons were predicted in 2D \cite{Malomed2001,Dimitri} and 3D
\cite{Panos3D} fully discrete media, including stable 2D modes with $S\geq 2$
\cite{Chen}. Such 2D discrete states were created in photorefractive
lattices \cite{Neshev,Segev}. The robustness of semidiscrete QDs presented
below, and available techniques for the work with QDs \cite{Cabrera2017}-%
\cite{Ing2} suggest that the creation of the semidiscrete states is a
relevant objective for experiments.

The setting is realized in the form of the system of linearly-coupled GPEs
for the semidiscrete wave function, $\psi _{j}(z,t)$ (in the basic form,
the same for both components of the binary condensate \cite%
{Petrov2015,Petrov2016}), with longitudinal coordinate $z$ and the
transverse discrete one, $j$. The GPE system includes the cubic
self-repulsion competing with the LHY-induced quadratic self-attraction. In
a scaled form \cite{Petrov2016,Astrakharchik2018}, it is
\begin{gather}
i\partial _{t}\psi _{j}=-{(1/2)}\partial _{zz}\psi _{j}-\left( C/2\right)
\left( \psi _{j+1}-2\psi _{j}+\psi _{j-1}\right)  \notag \\
+g|\psi _{j}|^{2}\psi _{j}-|\psi _{j}|\psi _{j}+\left( \omega ^{2}/2\right)
z^{2}\psi _{j},  \label{GPE}
\end{gather}%
where $C>0$ is the coupling between adjacent cores, the strength of the
quadratic attraction is normalized to be $1$, and $g>0$ is the strength of
the cubic self-repulsion. A realistic model should include a trapping
potential with strength $\omega ^{2}$ (its action in the discrete direction
is negligible, as the trapping effect of the OL potential, which makes the
setting semidiscrete, is much stronger). Estimates for physical parameters
of the system and predicted semidiscrete modes are given below.

It is relevant to mention studies of fully discrete 1D and 2D solitons \cite%
{CQ1D,Chong2009}, which are supported by the competition of cubic-quintic
onsite nonlinearities. Similarly, we find many branches of zero-vorticity
states, of onsite-centered (OC) and intersite-centered (IC) types, which are
chiefly stable, but tend to disappear with the increase of $C$, as the
medium is approaching a quasi-continuum\ (QC)\ regime. However, in the
present work we address semidiscrete modes, rather than fully discrete
ones, and we address semidiscrete vortices, with winding numbers $1\leq
S\leq 5$, that were not considered previously.

The total norm, $N=\sum_{j}\int_{-\infty }^{+\infty }\left\vert \psi
_{j}(z)\right\vert ^{2}dz$, is fixed by choosing its particular value.
First, for states with vorticities $S=0$ and $1$, the fixed value is $%
N_{S=0,1}=100$, which is appropriate for plotting the results [as shown
below, typical values of the actual (unscaled) number of atoms are $\sim
10^{4}$]. For $S\geq 2$, it is convenient to fix larger values of $N$. Two
remaining control parameters are $C$ and $g$, that will be varied in the
range of $0\leq C,g\leq 1$, which is sufficient for identifying all species
of self-trapped states and making conclusions about their stability. Along
with the norm, the system conserves the energy,%
\begin{gather}
E=\sum\nolimits_{j}\int_{-\infty }^{+\infty }\left\{C\left[ \left\vert \psi
_{j}\right\vert ^{2}-\frac{1}{2}\psi _{j}^{\ast }\left( \psi _{j+1}+\psi
_{j-1}\right) \right] \right.  \notag \\
\left. +\frac{1}{2}\left\vert\left( \psi _{j}\right) _{z}\right\vert ^{2}+%
\frac{g}{2}\left\vert \psi _{j}\right\vert ^{4}-\frac{2}{3}\left\vert \psi
_{j}\right\vert ^{3}+\frac{\omega ^{2}}{2}\left\vert \psi _{j}\right\vert
^{2}\right\} dz  \label{E}
\end{gather}

Stationary states with chemical potential $\mu $ are looked for as $\psi
_{j}(z,t)=\phi _{j}(z)e^{-i\mu t}$, where $\phi _{j}(z)$ is a localized wave
function. In the limit of $C=0$, the uncoupled GPE (\ref{GPE}) gives rise to
1D QDs. Particularly, they assume the flat-top shape with a nearly constant
density, $\left\vert \phi \right\vert ^{2}=4/(9g)$, at $\mu $ close to $\mu
_{0}=4/9-2/(3\sqrt{g})$, at which the QD's width diverges \cite%
{Petrov2016,Astrakharchik2018}. On the other hand, in the limit of $%
g\rightarrow 0$ the QD takes a well-localized shape, $\phi
_{g=0}(z)=(3/2)|\mu _{g=0}|\mathrm{sech}^{2}\left( \sqrt{|\mu _{g=0}|/2}%
z\right) $, with $\mu _{g=0}=-(1/3)\left( 2N^{2}/3\right) ^{1/3}$.

The anisotropy of 2D QDs in the $\left( z,j\right) $ plane is defined as the
ratio of its widths in the $z$ and $j$ directions, $\varepsilon ={\sqrt{C}%
L_{z}}/{L_{j}}$, with $L_{z}\equiv \left[ {\int_{-\infty }^{+\infty }|\phi
_{j=0}(z)|^{4}dz}\right] ^{-1}\left( \int_{-\infty }^{+\infty }|\phi
_{j=0}(z)|^{2}dz\right) ^{2}$ and $L_{j}\equiv \left[ {\sum_{j}|\phi
_{j}(z=0)|^{4}}\right] ^{-1}\left[ \sum_{j}|\phi _{j}(z=0)|^{2}\right] ^{2}$%
. Indeed, in the continuum limit ($C\rightarrow \infty $), $\varepsilon =1$
implies that the 2D mode is axially symmetric in the plane of coordinates $%
\left( z,j/\sqrt{C}\right) $. It is shown below that $\varepsilon $
determines a boundary between tightly-bound (TB) and QC semidiscrete states.

To generate stationary modes, Eq. (\ref{GPE}) was solved numerically, using
the imaginary-time and squared-operator \cite{Taras} methods for finding QDs
with $S=0$ and $S\geq 1$, respectively. Stability of the stationary states
was then identified through computation of eigenvalues for small
perturbations, and by dint of simulations of Eq. (\ref{GPE}) in real time,
both approaches producing almost identical results. The use of the GPE with
the LHY term is relevant for exploring stability of states supported by
quantum fluctuations, as implied by the derivation of the model \cite%
{Petrov2015,Petrov2016}, and confirmed by direct comparison of the
predictions with experimentally observed dynamics \cite%
{Cabrera2017,Cheiney2018,Ing1}.



\begin{figure}[t]
{\includegraphics[width=1.0\columnwidth]{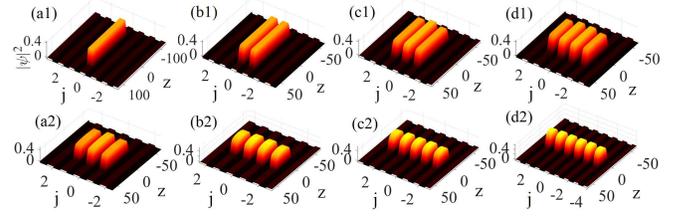}}
\caption{Density profiles of coexisting stable OC and IC semidiscrete QDs
with zero vorticity and the number of sites from $1$ to $4$ in panels
(a1-d1), respectively. Parameters are $\protect\omega =0$ and $%
(g,C,N)=(1,0.01,100)$. (a2-d2) QDs produced from those displayed in (a1-d1)
by the squeezing transition imposed by the trap with $\protect\omega %
=0.004,0.011,0.015$ and $0.021$ in Eq. (\ref{GPE}), respectively. The corresponding transition
points are $\protect\omega _{c}=0.003$, $0.010$, $0.014$, and $0.020$, and
the ratio of the QD's longitudinal size to the trapping length, $L_{\protect%
\omega }=\protect\omega _{c}^{-1/2}$, is $L_{z}/L_{\protect\omega }=10.20$, $%
8.11$, $6.74$, and $5.93$. }
\label{Fundamental}
\end{figure}

\begin{figure}[t]
{\includegraphics[width=1.0\columnwidth]{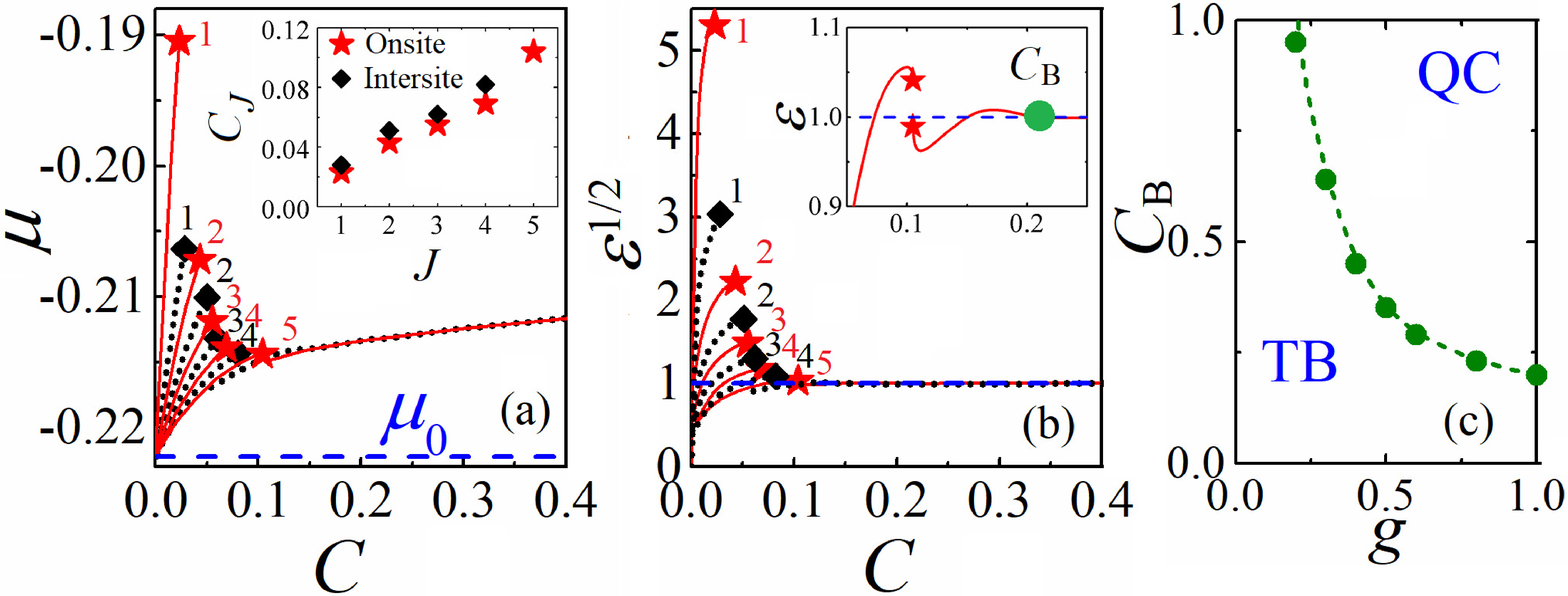}}
\caption{(a,b) Dependences $\protect\mu (C)$ and $\protect\sqrt{\protect%
\varepsilon (C)}$ for the OC and IC (red solid and black dotted lines,
respectively) QDs with zero vorticity, built of $2J-1$ or $2J$ sites, versus
$J$, at $g=1$ and $\protect\omega =0$. The dashed line in (a) is $\protect%
\mu _{0}(g=1)\approx -0.222$. These branches exist up to terminal values of $%
J$, \textit{viz}., $J_{\mathrm{T}}^{\mathrm{(on,in)}}=5$ and $4$,
respectively. The subplot in (a) shows critical values, $C_{J}$, at which
the respective branches terminate, vs. $J$. The subplot in (b) displays the
switch of the single eventually surviving onsite-centered mode from $J=5$ to
$6$ with the subsequent increase of $C$. The green point, $C_{\mathrm{B}}$,
at which $\protect\varepsilon =1$, is the boundary between the QDs with the
TB and QC structure, for $g=1$. (c) The TB-QC boundary, $C_{\mathrm{B}}(g)$.}
\label{character}
\end{figure}

\textit{Zero-vorticity QDs}. Two kinds of zero-vorticity modes, \textit{viz}%
., OC and IC ones, which occupy, respectively, $2J-1$ and $2J$ sites, are
produced by the numerical solution. Starting from the anti-continuum limit ($%
C\rightarrow 0$) \cite{Aubry}, many coexisting solutions are found,
corresponding to $1\leq J\leq 5$ and $1\leq J\leq 4$ for the OC and IC
configurations, respectively, i.e., with the number of sites from $1$ to $9$%
, see examples of stable semidiscrete modes with $J=1,2$ in Fig. \ref%
{Fundamental}. The coexisting solution branches are represented by the
respective dependences $\mu (C)$ and $\varepsilon (C)$, for the
above-mentioned fixed norm, $N_{S=0,1}=100$, and fixed $g$, in Fig. \ref%
{character}(a,b). The comparison of energy (\ref{E}) for different modes
demonstrates that the ground state (energy minimum) always corresponds to
the largest number of sites. For these states, the energy of the intersite
coupling is $\symbol{126}~2\%-5\%$ of the total energy, and $6\%-8\%$ for
the vortex states considered below.

The branches originate from $\mu =\mu _{0}$ and $\varepsilon =0$ at $C=0$,
and terminate at critical values $C=C_{J}$ \cite{Infact}, which are denoted
in Figs. \ref{character}(a) and (b) by red stars are black rhombuses for the
OC and IC states, respectively. The subplot in Fig. \ref{character}(a) shows
$C_{J}$ is a function of $J$, demonstrating that the multiplicity of
coexisting branches reduces, step by step, with the increase of $C$. The
single branch survives at $C\geq 0.10$, carrying over into the single
fundamental mode with $\varepsilon =1$ in the 2D continuum, as seen in Fig. %
\ref{character}(b). With the increase of $C$, the evolution of the single
surviving state proceeds through the increase of the number of sites in this
state. An example is the transition from $9$ to $11$ sites, with an
incremental increase of $C$, as shown in the inset to Fig. \ref{character}%
(b).


Semidiscrete QDs can be categorized as TB or QC ones, if their shapes
feature strong or weak discreteness, respectively, the boundary between them
being determined by proximity of $\varepsilon $ to $1$. The transition from $%
\varepsilon \neq 1$ to $\varepsilon =1$ is illustrated by the inset to Fig. %
\ref{character}(b), where $C_{\mathrm{B}}$ is identified as the transition
point. A boundary between the TB and QC regimes in the $\left( g,C\right) $
plane is displayed in Fig. \ref{character}(c). The decrease of the boundary
value, $C_{\mathrm{B}}$, with the increase of $g$ is a natural trend, as
stronger self-repulsion makes the droplet broader in the discrete direction,
which is the same effect as produced by stronger coupling.

\begin{figure}[t]
{\includegraphics[width=1.0\columnwidth]{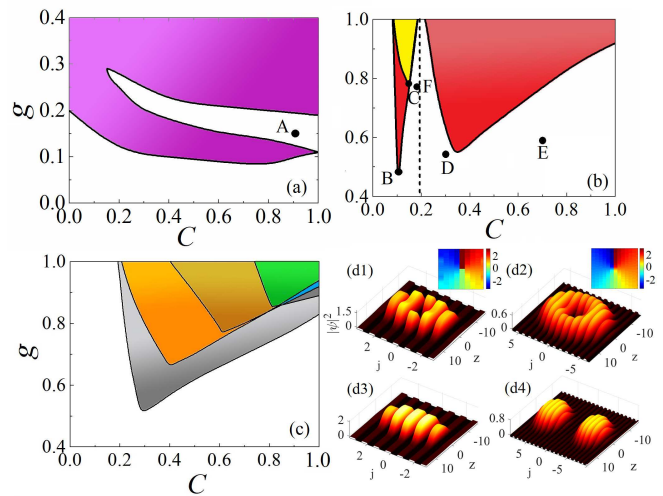}}
\caption{(a) The stability area (purple) for IC semidiscrete QDs with zero
vorticity, in the $(C,g)$ parameter plane. Unstable modes, such as the one
corresponding to point A, with $(g,C)=(0.9,0.15)$, transform into robust
breathers, as shown in Fig. \protect\ref{unstable}(a). Panel (b) displays
stability areas for OC and IC vortex QDs with $S=1$ (yellow and red regions,
and the yellow-only one, respectively). In panels (a) and (b), the fixed
normalization is $N=100$. (c) Stability areas for OC vortices with $S=2$
(all colored regions), $3$ (orange + brown + green), $4$ (brown + green +
blue), and $5$ (green + blue + dark gray). For the convenience of plotting,
the normalization for $S=2$ through $5$ is fixed as $N=400,900,2500$ and $%
4500$, respectively. (d1,d2) Examples of stable onsite- and
intersite-centered vortices, for $(g,C)=(0.48,0.1)$ and $(0.77,0.15)$, which
correspond, respectively, to points B and C in panel (c). The insets display
the respective phase profiles. (d3,d4) Fundamental and dipole-mode QDs, into
which the squeezing transforms the onsite- and intersite-centered vortices
at $\protect\omega =0.024$ and $0.006$. The corresponding critical values
are $\protect\omega _{c}^{\prime }=0.023$ and $0.005$, respectively.}
\label{stable}
\end{figure}

\begin{figure}[t]
{\includegraphics[width=1\columnwidth]{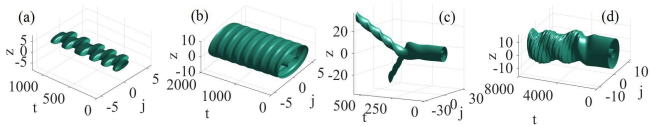}}
\caption{Examples of the evolution of unstable semidiscrete QDs. (a) The OC
QD with $S=0$, denoted by point A in Fig. \protect\ref{stable}(a),
transforms into a breather which performs shuttle motion. (b,c,d) Vortex QDs
with $(g,C)=(0.55,0.3)$, $(0.65,0.6)$ and $(0.75,0.18)$, which are denoted
by points D-F in Fig. \protect\ref{stable}(d).}
\label{unstable}
\end{figure}

Analysis of the stability of the semidiscrete QDs with $S=0$ demonstrates
that the OC modes are stable in the entire $(g,C)$ plane, while their IC
counterparts are stable only in a part of the plane [Fig. \ref{stable}(a)].
The instability of the latter states at small $g$, and their stabilization
at larger $g$, is similar to findings for 1D discrete solitons in the model
with the cubic-quintic nonlinearity \cite{CQ1D}. However, a new feature is
an inner lacuna in the stability area. In direct simulations, unstable IC
QDs transform into robust breathers, which perform shuttle oscillations
[Fig. \ref{unstable}(a)].

With the increase of $\omega $ (the trap's strength), QDs found at $\omega =0
$ undergo a \textit{squeezing transition} at critical values $\omega _{c}$,
which transforms them into stable QDs with two sites added at their edges,
as shown in Fig. \ref{Fundamental}(a2-d2). For QDs with larger numbers of
sites, \textit{viz}., $5,6,7,8$, the
respective critical values are $\omega _{c}=0.025,0.038,0.032,0.045$ and
$L_{z}/L_{\omega }=5.33,4.94,4.62,4.37$.
The large values of the length ratio imply robustness of the QDs
against the squeezing.

\textit{Vortex modes}. Semidiscrete states with vorticity $S$ represent a
novel species of self-trapped modes, their stability being a central issue,
as suggested by studies of vortex solitons in continuous models \cite{PhysD}%
. The present system gives rise to such states at $g\geq g_{\min }\approx $ $%
0.4$. They seem as quasi-isotropic modes, with $\varepsilon $ close to $1$.
The systematic numerical analysis identifies stability areas for OC vortex
QDs with $S=1$ and $2\leq S\leq 5$, which are displayed in Figs. \ref{stable}%
(b) and (c), respectively. For $S=1$, the stability area is split in two
parts at $C\approx 0.19$ [by the dashed line in Fig. \ref{stable})(b)],
approximately equal to value $C_{\mathrm{B}}$ separating the TB and QC
regions at $g=1$ in Fig. \ref{character}(d). Vortices with $S\geq 2$ were
found only at $C>0.19$.

There are two different kinds of stable vortices with $S=1$, of the OC and
IC types, with pivots located, respectively at a lattice site or between two
sites. Stable IC vortices are found only in a small yellow parameter region
in Fig. \ref{stable}(b) at $C<0.19$, where the semidiscrete states feature
a TB structure, and they do not exist with $S\geq 2$ (in fully discrete 2D
lattices it is also difficult to find stable IC vortex solitons \cite%
{Chong2009}). Stable OC and IC vortex QDs [points B and C in Fig. \ref%
{stable}(a)] are displayed in Fig. \ref{stable}(d1,d2). In the QC region ($%
C>0.19$), stability areas for OC vortices with $2\leq S\leq 5$ are displayed
in Fig. \ref{stable}(c), being similar to their counterpart with $S=1$ in
Fig. \ref{stable}(b). The vortex QDs are stable at values of $g$ exceeding a
certain threshold, which gradually increases with $S$.

With the increase of the trap's strength, $\omega $, the squeezing
transition leads to destabilization of OC and IC vortices from Figs. \ref%
{stable}(d1,d2) at $\omega _{c}=0.005$ and $0.001$, respectively, the
corresponding size ratios being $L_{z}/L_{\omega }=0.94$ and $0.60$, i.e.,
the vortices are more fragile states than the fundamental states. With\ the
further increase of $\omega $, the unstable vortices are replaced by stable
fundamental and dipole-mode QDs at $\omega _{c}^{\prime }=0.023$ and $0.005$
[Fig. \ref{stable}(d3,d4)], the corresponding length ratios being $%
L_{z}/L_{\omega }\simeq 1.83$ and $1.42$.


Unstable vortex QDs display different evolution scenarios. Close to the
stability boundary, they form robust breathers which keep initial $S$ [Fig. %
\ref{unstable}(b)]. Far from the stability boundary, an unstable vortex QD
splits in two fragments [Fig. \ref{unstable}(c)]. Near the TB-QC boundary,
unstable vortices undergo conspicuous deformation, but do not split, keeping
$S$ and featuring chaotic evolution in Fig. \ref{unstable}(d). Its chaotic
character is confirmed, following a known criterion \cite{chaos}, by
computation of the power spectrum of oscillations of the peak density, in
which $28\%$ of the total power belongs to a continuous component.

Undoing the rescaling used in the derivation of Eq. (\ref{GPE}) \cite%
{Petrov2015,Petrov2016}, we conclude that a longitudinal size of the states
is expected to be $L_{z}\sim 10$ $\mathrm{\mu }$m with $\symbol{126}%
10^{4}-10^{5}$ atoms of $^{39}$K \cite{Cabrera2017}-\cite{Ing2}, transverse
trap $\omega _{\mathrm{tr}}\sim 2\pi \times 200$ Hz, and the OL potential
with wavelength $\sim 4$ $\mathrm{\mu }$m and respective recoil energy, $2m_{%
\mathrm{atom}}^{-1}\left( \pi \hbar /\lambda \right) ^{2}$. The coupling
constant, $C$, may be adjusted by variation of the OL depth \cite{Smerzi}%
.The critical strength of the longitudinal trap, which initiates the
squeezing transition, is $\sim \omega _{\mathrm{tr}}$ for the robust $S=0$
states, while for more fragile vortices it is $\sim 2\pi \times 20$ Hz.
Vorticity may be imparted to the condensate by a helical optical beam,
transversely focused on spotsize $\sim L_{z}$ \cite{Davidson}. The
experimental realization is definitely possible at temperatures $\lesssim
2.5 $ $\mathrm{\mu }$K \cite{Cheiney2018}. Due to three-body losses, the
modes will start to decay at $t\gtrsim 50$ ms, which allows one to observe
them by means of available techniques \cite{Cabrera2017,Cheiney2018,Ing1}.

\textit{Conclusion}. We have introduced a setting for the study of
semidiscrete QDs in the form of the array of 1D guides coupled by hopping
of atoms. Each guide is filled by a binary condensate, which gives rise to a
semidiscrete system, in the form of the GPE including the repulsive cubic
and attractive quadratic (LHY)\ terms, with the longitudinal continuous and
transverse discrete coordinates. The systematic analysis reveals many
families of stable 2D semidiscrete QDs, of the onsite- and
intersite-centered (OC and IC) types, which terminate one by one with the
increase of the coupling coefficient. The system's parameter space splits
into tight-binding (TB) and quasi-continuum (QC) parts, with a single stable
family surviving in the latter one. Previously unexplored self-trapped modes
are semidiscrete vortices. In the TB region, vortex QDs, of both OC and IC
types, are stable with winding number $S=1$, while in the QC region OC
vortices remain stable up to $S=5$. The application of the longitudinal trap
leads to squeezing transitions of $S=0$ states, and, eventually, to
transformation of vortices into fundamental or dipole modes.

A similar setting may be implemented for spatial optical solitons in stacks
of planar waveguides with cubic-quintic nonlinearity \cite{Cid2}. A
challenging extension is to consider a 3D setting with two discrete
coordinates.

\begin{acknowledgments}
This work was supported by NNSFC (China) through Grant No. 11874112, No. 11575063 and No. 11905032, Guangdong Provincial Department of Education Project through Grant No. 2018KQNCX279, the Israel Science Foundation through Grant No. 1286/17, and by the special Funds for the Cultivation of Guangdong College students Scientific and Technological innovation No. pdjh2019b0514
\end{acknowledgments}


\begin{thebibliography}{99}
\bibitem{Infact} In fact, the branches terminate at the critical points through saddle-note bifurcations generated by collision with additional branches of unstable solutions, which are not shown in the figure.
\end{thebibliography}
\end{document}